\documentclass{article}
\usepackage{graphicx}
\usepackage{booktabs}
\usepackage{enumitem}
\usepackage{tabulary}

\usepackage[T1]{fontenc}
\usepackage[utf8]{inputenc}

\usepackage{arxiv_style}
\usepackage{amsmath,cite,url, amsfonts, microtype, multirow}
\usepackage{graphicx}
\usepackage{color}
\usepackage{enumitem}
\usepackage{tikz}
\usetikzlibrary{calc}
\usepackage{adjustbox}
\usepackage{stfloats}
\usepackage[colorlinks=false,linkcolor=blue,citecolor=blue,bookmarks=false]{hyperref}
\oneauthor
  {Scott H. Hawley\thanks{Email: scott.hawley@belmont.edu}}
  {Department of Chemistry \& Physics, Belmont University, Nashville, TN, USA}

\title{MIDI-RAE-JEPA: Hierarchical Representation Learning and Generation for Symbolic Music}

\begin{document}

\maketitle

\begin{abstract}

Rich internal representations of musical structure are essential for music understanding tasks such as machine-assisted music co-writing, yet self-supervised approaches for symbolic music representation remain underexplored, particularly those that encode the hierarchical multiscale nature of musical structures. We present MIDI-RAE-JEPA, combining a pitch- and time-shift equivariance objective with LeJEPA and a Swin Transformer V2 encoder to learn such hierarchical representations of symbolic music encoded as piano roll images. The time-shift equivariance objective encourages the model to internalize temporal musical relationships. The encoder is trained purely on self-supervised objectives --- including a masked embedding predictor (MEP) --- with collapse prevented via SIGReg. A separate decoder trained on the frozen encoder embeddings achieves reconstruction F1 of 0.995, and a flow matching generative model conditioned on those embeddings produces generations that closely match the pitch register and rhythmic density of the conditioning excerpt, while mismatched conditioning yields unrelated but musically plausible output. Learned representations outperform a Haar scattering transform baseline on a downstream emotion classification task, and embedding distances increase monotonically with pitch and time shift magnitude, confirming measurable equivariance. These results suggest that equivariance-based SSL objectives, combined with sufficient fine-level encoder capacity, provide a viable path toward semantically rich, generatively useful representations of symbolic music.

\end{abstract}

\section{Introduction}

A long-standing goal in music technology is an intelligent AI assistant that can listen to a musical idea from a human and provide meaningful feedback to refine the composition, arrangement, or production.
Such a system requires rich internal representations of musical structure that capture not only local note patterns but also the hierarchical relationships---notes forming short phrases, phrases combining into sections, sections shaping entire pieces---that characterize how music is organized~\cite{lerdahl1983generative}.
Large multimodal language models, despite their breadth, currently perform poorly at music understanding tasks~\cite{ma2025cmibench}, motivating domain-specific approaches.

Symbolic music encoded as piano rolls presents sharply different structure from natural images.
A piano roll is a binary $128 \times 128$ image where the horizontal axis encodes time (eighth-note resolution over 8 bars) and the vertical axis encodes pitch (all 128 MIDI pitches).
Piano roll representations have proven robust and effective for generative music modeling, as demonstrated by Polyffusion~\cite{min2023polyffusion} among others.
This controlled sparsity also makes piano rolls a useful testbed for methods intended for audio spectrograms: the two representations share a time-frequency grid structure, but piano rolls strip away acoustic complexity, allowing architectural and objective choices to be evaluated in isolation before scaling to the messier audio domain.

A natural SSL objective for images is to have embeddings of crops predict the embedding of the full image, as in DINOv2~\cite{dinov2}.
However, piano rolls are far out of distribution from the natural photographs on which such models are trained, and the crop-and-resize strategy central to DINOv2 is semantically problematic: ``zooming in'' on a piano roll would merge adjacent semitones and distort rhythmic structure, unlike the smooth scale invariance of photographs.
We therefore adopt an alternative SSL objective based on equivariance to pitch and time translation---musically natural transformations that preserve structure while shifting position---combined with a masked embedding predictor in the spirit of JEPA~\cite{assran2023ijepa}.

Recently, Representation Autoencoders (RAEs)~\cite{zheng2025rae} showed that rich pretrained representations can support not only discriminative tasks but also high-quality generation when paired with a separately trained decoder.
A natural question is whether this paradigm extends to music.
We show that it does: our encoder representations support reconstruction, conditional generation, and downstream classification.

We make the following contributions:

(1) We present a system for \textbf{hierarchical SSL representations for symbolic music}, combining equivariance objectives, LeJEPA, and a Swin V2 encoder; to our knowledge the first hierarchical JEPA-based approach for this domain.

(2) We demonstrate a \textbf{complete RAE pipeline for music} --- encoder, frozen-encoder decoder, and conditioned generative model --- showing the representations are useful end-to-end for both reconstruction and generation.

(3) We provide \textbf{empirical validation}: equivariance holds measurably in pitch and time; representations transfer to a downstream emotion classification task; and conditioning controls generation meaningfully.

(4) We introduce a \textbf{smooth pitch- and time-translation equivariance loss} with a delta-scaled target distance that both attracts and repels embedding pairs, preventing collapse while maintaining a geometrically consistent latent space.

(5) We introduce a \textbf{soft factorization loss} that encourages pitch and time directions to be geometrically orthogonal in latent space.

All experiments run on consumer hardware (RTX 4090, 4090 MaxQ, RTX 2070), demonstrating feasibility without large-scale compute. Code is available on GitHub.\footnote{\url{https://github.com/drscotthawley/midi-rae}}

\subsection{Related Work}

\textbf{Joint-Embedding Predictive Architectures.}
LeJEPA~\cite{balestriero2025lejepa} provides a theoretically grounded
self-supervised objective combining an attraction loss with Sketched Isotropic
Gaussian Regularization (SIGReg), requiring no stop-gradients, no negative
pairs, and no complex training tricks.
I-JEPA~\cite{assran2023ijepa} and DINO~\cite{caron2021dino} provide related
JEPA and teacher-student frameworks we draw on.
JEPA-style objectives have been applied to various data formats. M2D~\cite{niizumi2023m2d}
applies masked modeling with dual networks to learn audio representations. V-JEPA 2~\cite{assran2025vjepa2selfsupervisedvideo} applies JEPA principles for video understanding.
To our knowledge, no prior work applies JEPA-style SSL to symbolic music
representations, hierarchical or otherwise.

\textbf{Representation Autoencoders.}
The RAE paradigm~\cite{zheng2025rae} replaces VAE compression with frozen pretrained encoders paired with trained decoders, operating directly in semantically structured representation space.
We adopt this spirit for music: the encoder learns representations suitable for both discriminative tasks and downstream generation.

\textbf{Equivariant SSL for Music.}
Multiple recent works apply equivariance-based self-supervised learning to audio.
PESTO~\cite{riou2023pesto} estimates pitch using a Siamese network with a
Toeplitz output layer that enforces hard pitch-equivariance.
Quinton~\cite{quinton2022equivariant} applies equivariant self-supervision to
tempo estimation, treating time-stretching as the invariant transformation.
STONE~\cite{kong2024stone} learns tonality representations equivariant to
pitch transposition.
Cwitkowitz and Duan~\cite{cwitkowitz2024ssmpe} pursue fully self-supervised
multi-pitch estimation 
via equivariance to geometric transformations and invariance to timbral variations.
Our work addresses a related but distinct problem: 2D equivariance to
\emph{simultaneous} pitch and time translation in the symbolic domain,
learned softly through a metric objective in a hierarchical encoder rather
than enforced architecturally.

\textbf{Symbolic Music Representation Learning.}
Transformer-based pretraining on symbolic music event sequences has produced
strong representations for classification tasks.
MusicBERT~\cite{zeng2021musicbert} adapts BERT-style masked token prediction
to OctupleMIDI-encoded sequences.
MidiBERT-Piano~\cite{chou2024midibert} pretrains on piano performance MIDI
and fine-tunes on several downstream tasks.
MuseBERT~\cite{wang2021musebert} extends this paradigm to controllable
generation.
These sequence-based approaches operate on event tokens and rely on
fine-tuning for downstream tasks; our approach instead operates on piano-roll images, learning to 
predict notes outside the receptive field of the visible image crop 
and yielding representations that transfer without fine-tuning via 
linear probes.

\section{Method}

\subsection{Data Representation}

We use POP909~\cite{wang2020pop909}, 909 popular songs in MIDI format with separated melody and accompaniment.
Each 8-bar segment is rendered as a $128\times128$ binary piano roll (union of melody and accompaniment), with 64 eighth-note time steps horizontally and all 128 MIDI pitches vertically.
While larger MIDI corpora exist (e.g., the Lakh dataset~\cite{raffel2016lakh}), POP909 serves as an effective sandbox: small enough for rapid iteration, varied enough to avoid overfitting, and well-curated for structural analysis.
Random crop sampling with pitch and time transposition augmentation means the model rarely encounters an identical input twice, greatly expanding the effective training variety available from a compact dataset.

\subsection{Hierarchical Encoder}

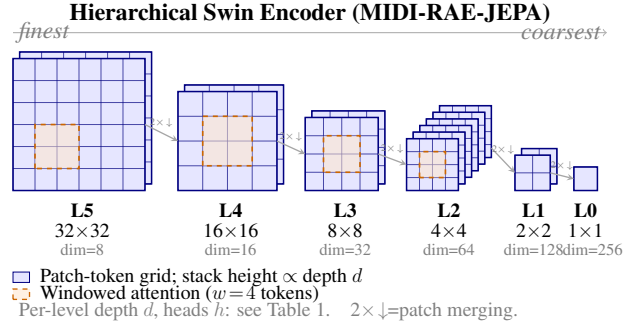
\begin{figure}[tbp]
  \centering
  \resizebox{\columnwidth}{!}{\begin{tikzpicture}[font=\small]

\colorlet{gridface}{blue!10}
\colorlet{gridborder}{blue!50!black}
\tikzset{
  grid/.style={fill=gridface, draw=gridborder, line width=0.6pt},
  wbox/.style={draw=orange!80!black, line width=1.0pt, dashed,
               fill=orange!20, fill opacity=0.55},
  dnarrow/.style={->, >=stealth, thick, gray!70},
}

\def\OX{0.14}  
\def\OY{0.14}  

\newcommand{\drawgrid}[4]{%
  \pgfmathsetmacro{\cs}{(#3)/(#4)}
  \fill[gridface] (#1,#2) rectangle (#1+#3, #2+#3);
  \foreach \i in {0,...,#4} {
    \draw[gridborder, line width=0.4pt, opacity=0.7]
      (#1+\i*\cs, #2) -- (#1+\i*\cs, #2+#3);
    \draw[gridborder, line width=0.4pt, opacity=0.7]
      (#1, #2+\i*\cs) -- (#1+#3, #2+\i*\cs);
  }
  \draw[gridborder, line width=0.9pt] (#1,#2) rectangle (#1+#3,#2+#3);
}

\newcommand{\drawstack}[5]{%
  \foreach \li in {1,...,#5} {
    \pgfmathsetmacro{\lb}{#5 - \li}   
    \pgfmathsetmacro{\px}{#1 + \lb*\OX}
    \pgfmathsetmacro{\py}{#2 + \lb*\OY}
    \drawgrid{\px}{\py}{#3}{#4}
  }
  \pgfmathparse{int(#4 >= 3)}
  \ifnum\pgfmathresult=1
    \pgfmathsetmacro{\cs}{(#3)/(#4)}
    \draw[wbox] (#1+\cs, #2+\cs) rectangle (#1+3*\cs, #2+3*\cs);
  \fi
}

%
%

\begin{scope}[yshift=-0.35cm]
\def\BY{0}

\drawstack{0}{\BY}{2.8}{6}{2}

\drawstack{3.5}{\BY}{2.1}{4}{2}

\drawstack{6.2}{\BY}{1.55}{4}{2}

\drawstack{8.35}{\BY}{1.1}{4}{6}

\drawstack{10.65}{\BY}{0.75}{2}{2}

\drawstack{11.9}{\BY}{0.50}{1}{1}

\draw[dnarrow] (2.8, 1.4)   -- (3.5, 1.05)
  node[midway, above, font=\scriptsize, gray] {2$\times\!\downarrow$};
\draw[dnarrow] (5.6, 1.05)  -- (6.2, 0.775)
  node[midway, above, font=\scriptsize, gray] {2$\times\!\downarrow$};
\draw[dnarrow] (7.75, 0.775) -- (8.35, 0.55)
  node[midway, above, font=\scriptsize, gray] {2$\times\!\downarrow$};
\draw[dnarrow] (9.45+5*\OX, 0.55+5*\OY*0.5)
  -- (10.65, 0.375)
  node[midway, above, font=\scriptsize, gray] {2$\times\!\downarrow$};
\draw[dnarrow] (11.4, 0.375) -- (11.9, 0.25)
  node[midway, above, font=\scriptsize, gray] {2$\times\!\downarrow$};

\def\LY{-0.42}   
\def\SY{-0.86}   
\def\MY{-1.24}   

\node[font=\large\bfseries]     at (1.47, \LY) {L5};
\node[font=\large]          at (1.47, \SY) {32$\times$32};
\node[font=\normalsize, gray] at (1.47, \MY) {dim=8};

\node[font=\large\bfseries]     at (4.62, \LY) {L4};
\node[font=\large]          at (4.62, \SY) {16$\times$16};
\node[font=\normalsize, gray] at (4.62, \MY) {dim=16};

\node[font=\large\bfseries]     at (7.05, \LY) {L3};
\node[font=\large]          at (7.05, \SY) {8$\times$8};
\node[font=\normalsize, gray] at (7.05, \MY) {dim=32};

\node[font=\large\bfseries]     at (9.25, \LY) {L2};
\node[font=\large]          at (9.25, \SY) {4$\times$4};
\node[font=\normalsize, gray] at (9.25, \MY) {dim=64};

\node[font=\large\bfseries]     at (11.05, \LY) {L1};
\node[font=\large]          at (11.05, \SY) {2$\times$2};
\node[font=\normalsize, gray] at (11.05, \MY) {dim=128};

\node[font=\large\bfseries]     at (12.15, \LY) {L0};
\node[font=\large]          at (12.15, \SY) {1$\times$1};
\node[font=\normalsize, gray] at (12.15+0.15, \MY) {dim=256};
\end{scope}

\node[font=\Large\bfseries] at (6.3, 3.4)
  {Hierarchical Swin Encoder (MIDI-RAE-JEPA)};

\draw[->, gray!60, line width=0.7pt] (0, 3.0) -- (12.6, 3.0);
\node[font=\Large, gray, anchor=west]  at (0,   3.0) {\textit{finest}};
\node[font=\Large, gray, anchor=east]  at (12.6,3.0) {\textit{coarsest}};

\begin{scope}[shift={(0, -2.3)}]
  \fill[gridface] (0,0) rectangle (0.35, 0.22);
  \draw[gridborder, line width=0.6pt] (0,0) rectangle (0.35, 0.22);
  \node[anchor=west, font=\large] at (0.45, 0.11)
    {Patch-token grid; stack height $\propto$ depth $d$};

  \fill[orange!20, fill opacity=0.55]
    (0, -0.36) rectangle (0.35, -0.14);
  \draw[orange!80!black, dashed, line width=0.8pt]
    (0, -0.36) rectangle (0.35, -0.14);
  \node[anchor=west, font=\large] at (0.45, -0.25)
    {Windowed attention ($w\!=\!4$ tokens)};

  \node[anchor=west, font=\large, gray] at (0, -0.62)
    {Per-level depth $d$, heads $h$: see Table~1.\quad
     2$\times\!\downarrow$=patch merging.};
\end{scope}

\end{tikzpicture}}
  \caption{Our hierarchical Swin V2 encoder architecture.}
  \label{fig:arch}
\end{figure}

We use Swin Transformer V2~\cite{liu2021swin,liu2022swinv2}, which processes images hierarchically using shifted windows for  self-attention among neighboring patches.
Each stage downsamples by $2\times$, doubling the embedding dimension.
Our architecture uses $4\times4$ pixel patches on $128\times128$ pixel input, yielding a $32\times32$ initial patch grid.
The hierarchy has 6 stages, each downsampling $2\times$, as detailed in Table \ref{tab:arch} and illustrated in Figure \ref{fig:arch}.
Window size is 4 throughout. 

\begin{table}[tbp]
\centering
\caption{Architecture configuration. Visualized in Figure~\ref{fig:arch}.}
\label{tab:arch}
\begin{tabulary}{\linewidth}{ccccCC}
\toprule
Level & Patch size & grid & depth & heads & dim \\
\midrule
5 & $4\times4$px & $32\times32$ & 2 & 2 & 8 \\
4 & $8\times8$ & $16\times16$ & 2 & 2 & 16 \\
3 & $16\times16$ & $8\times8$ & 2 & 2 & 32 \\
2 & $32\times32$ & $4\times4$ & 6 & 4 & 64 \\
1 & $64\times64$ & $2\times2$ & 2 & 8 & 128 \\
0 & $128\times128$ & $1\times1$ & 1 & 16 & 256 \\
\bottomrule
\end{tabulary}
\end{table}

\subsection{Encoder Training Objective}

The total training objective combines three terms:
$$\mathcal{L} = \lambda \mathcal{L}_{\text{equiv}} + (1-\lambda) \mathcal{L}_{\text{SIGReg}} + \lambda_{\text{MEP}} \mathcal{L}_{\text{MEP}}$$
applied independently at each hierarchy level (excluding the two finest levels
for $\mathcal{L}_{\text{equiv}}$ and $\mathcal{L}_{\text{SIGReg}}$, where patch-level
equivariance is less meaningful).

Model variants marked with $\wedge$ (defined below) add a fourth term,
$\lambda_{\text{fact}} \mathcal{L}_{\text{fact}}$, which is the Soft Factorization Loss, described below.


\textbf{View generation.}
Two views are created by applying random shifts in time and pitch:
$\mathbf{x}_1$ is the original crop and $\mathbf{x}_2$ is shifted by
$(\Delta_x, \Delta_y)$ pixels.
Both views are full $128\times128$ windows extracted from the song's complete
piano roll; a ``shift'' is implemented by sampling the second window at a
position offset by $(\Delta_x, \Delta_y)$, so views always retain the encoder's
input size.


\textbf{Equivariance loss.}
A naive attraction loss collapses all shifted pairs to identical embeddings
regardless of shift magnitude.
Instead we enforce a \emph{target} embedding distance proportional to the
shift magnitude, so that larger shifts produce proportionally more distant embeddings:
$$\mathcal{L}_{\text{equiv}}(\mathbf{z}_1, \mathbf{z}_2, \boldsymbol{\delta}) =  \left(\|\mathbf{z}_1 - \mathbf{z}_2\| - \alpha \sqrt{d} \, \|\hat{\boldsymbol{\delta}}\|\right)^2$$
where $\mathbf{z}_1$ is the teacher embedding of $\mathbf{x}_1$ and
$\mathbf{z}_2$ is the student embedding of $\mathbf{x}_2$ (see the EMA teacher
paragraph below), $\|\cdot\|$ denotes the Euclidean (L2) norm, $d$ is the
embedding dimension, $\hat{\boldsymbol{\delta}}$ is the shift
$\boldsymbol{\delta}=(\Delta_x,\Delta_y)$ with each component divided by its
per-axis maximum ($\Delta t_{\max}$, $\Delta p_{\max}$; defined below) so that
each component lies in $[0,1]$, and $\alpha$ is a scalar hyperparameter.
The $\sqrt{d}$ factor accounts for concentration of measure in high-dimensional
spaces, giving $\alpha$ consistent semantic meaning across all hierarchy levels.
Unlike a hinge loss, this smooth quadratic both attracts pairs that are too far
apart \emph{and} repels pairs that are too close, maintaining a geometrically
consistent latent space.

Equivariance training uses patches shifted in pitch and time up to ($\Delta t_{\max}$, $\Delta p_{\max}$) pixels, with each shift sampled from a $\mathrm{Beta}(2,2)$ distribution, which yields a slight overemphasis on small shifts while still covering the full range. For the measurements in this paper, the maximum pitch shift $\Delta p_{\max}$ is 12 pixels (one octave) and the maximum time shift $\Delta t_{\max}$ is either
12 or 48. We denote the base configuration $\Delta t_{\max}$=12 as MRJ-12. A longer-shift variant with $\Delta t_{\max}$=48 is MRJ-48, and MRJ-12$\wedge$ includes the additional factorization loss component described below.


\textbf{Chunked SIGReg.}
SIGReg uses the Epps-Pulley characteristic function test to enforce an isotropic Gaussian prior on embeddings~\cite{balestriero2025lejepa}, theoretically optimal for downstream tasks.
SIGReg computes a tensor of shape $(N, N_{\text{slices}}, T)$ where $N_{\text{slices}}=256$.
On 16\,GB GPUs this tensor alone consumed $\sim$5\,GB.
We chunk the slice dimension into groups of 32 and accumulate in float32:
$$\mathcal{L}_{\text{SIGReg}} \approx \textstyle\sum_{k} \mathcal{L}^{(k)}_{\text{SIGReg}}$$
where $k$ indexes the chunks of 32 slices. This reduces peak VRAM by $\sim$5\,GB and counterintuitively improves throughput
by reducing memory allocator pressure.
We apply SIGReg only to student embeddings $\mathbf{z}_2$, ensuring correct
gradient flow.


\textbf{EMA Teacher and Masked Embedding Predictor.}
Following DINOv2~\cite{caron2021dino} and I-JEPA~\cite{assran2023ijepa}, we
maintain an EMA teacher $\theta_T \leftarrow \eta\,\theta_T + (1-\eta)\,\theta_S$
with $\eta = 0.96$, initialised from the student.
The teacher serves two roles simultaneously: it produces stable target embeddings
$\mathbf{z}_1$ from $\mathbf{x}_1$ for the equivariance loss, and it processes
$\mathbf{x}_2$ without masking to supply prediction targets for the MEP.
A lightweight predictor network, conditioned on the student's full-context
representation, predicts the teacher embeddings at randomly masked patch positions.

We note that our EMA teacher is \emph{not} redundant with SIGReg: in LeJEPA,
removing the teacher is possible because SIGReg alone prevents collapse, and
that remains its sole role here. Our teacher instead exists to provide stable,
slowly-varying regression targets for the equivariance and MEP losses; the two
mechanisms address different problems.

The MEP loss is averaged over hierarchy levels:

$$\mathcal{L}_{\text{MEP}} = \frac{1}{L}\sum_{\ell=1}^{L}
    \left\|\hat{\mathbf{e}}_\ell - \mathbf{e}^T_\ell\right\|^2$$

where $\mathbf{e}^T_\ell$ are the teacher embeddings and $\hat{\mathbf{e}}_\ell$
are the predictor's estimates at masked positions.
This encourages the student to develop predictive representations of musical
context across the full hierarchy, beyond what the equivariance objective alone enforces.

\begin{figure}[tbp]
\centering
\includegraphics[width=\columnwidth]{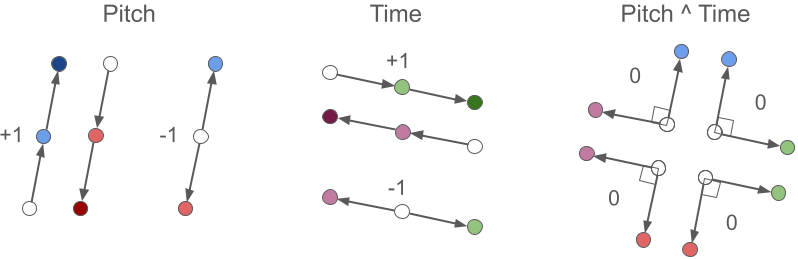}
\caption{Soft factorization targets. Differences between pairs of embeddings are encouraged to be parallel ($t{=}+1$), anti-parallel ($t{=}-1$), or orthogonal ($t{=}0$) depending on augmentation type and sign. Directions are not prescribed; only their pairwise geometry is constrained.}
\label{fig:softfact}
\end{figure}

\textbf{Soft factorization of pitch and time.}
Much of musical variety arises from interacting pitch and rhythmic patterns, suggesting that a latent space factored along pitch and time directions would support both structured generation and motif analysis.
Rather than enforcing factorization architecturally---which prior work has shown loses meaningful cross-dimension interactions~\cite{wu2023melodyglm,chen2020musicsketchnet}---we instead impose a soft geometric constraint through a cosine-similarity loss on augmentation difference vectors.
For a triplet of embeddings $\mathbf{z}_a$, $\mathbf{z}_1$, $\mathbf{z}_2$, where the anchor $\mathbf{z}_a$ is the embedding of the original unshifted crop and $\mathbf{z}_1$, $\mathbf{z}_2$ are embeddings of two independently shifted crops of the same excerpt, the factorization loss is:
$$
\begin{aligned}
\mathbf{d}_1 &= \mathbf{z}_1 - \mathbf{z}_a, \quad \mathbf{d}_2 = \mathbf{z}_2 - \mathbf{z}_a \\
\mathcal{L}_{\text{fact}} &= \left( \cos(\mathbf{d}_1, \mathbf{d}_2) - t \right)^2
\end{aligned}
$$
where the target $t \in \{+1, 0, -1\}$ encodes the geometric relationship dictated by the augmentation types: same-type same-sign shifts should be \emph{parallel} ($t{=}+1$), same-type opposite-sign shifts \emph{anti-parallel} ($t{=}-1$), and cross-type shifts (one pitch, one time) \emph{orthogonal} ($t{=}0$).
No directions are prescribed; the system discovers them freely subject to these pairwise constraints.
Figure~\ref{fig:softfact} illustrates the three target relationships.

\subsection{Decoder}

The decoder inverts the Swin encoder hierarchy: a mirrored stack of transformer stages with increasing spatial resolution, coupled with a Feature Pyramid Network (FPN) that propagates information from coarse levels down to the finest level via cross-attention-like skip connections.
The decoder is trained with embeddings from the \emph{frozen} pretrained encoder.
The reconstruction objective for binary piano roll images is binary cross entropy per pixel, with a small amount of label smoothing and an auxiliary MSE term to aid early spatial registration.
Decoded outputs are binarized at a threshold of 0.5.

\subsection{Generative Model}

We train a flow matching generative model~\cite{lipman2023flow} in pixel space, conditioned on encoder representations.
Flows operate between noise and data using the standard linear path $x_t = (1-t)x_0 + t x_1$, with velocity learned by a U-Net.
Source--target pairs are ordered per mini-batch using exact optimal transport (OT) following~\cite{tong2024improving}, using Tong et al.'s implementation directly.\footnote{\url{https://github.com/atong01/conditional-flow-matching}}
Exact OT pairing accelerates training convergence and provides a near-guarantee that conditioning signals applied during training align with the correct destination marginal of the fully trained flow.

\begin{figure}[tbp]
\centering
\includegraphics[width=\columnwidth]{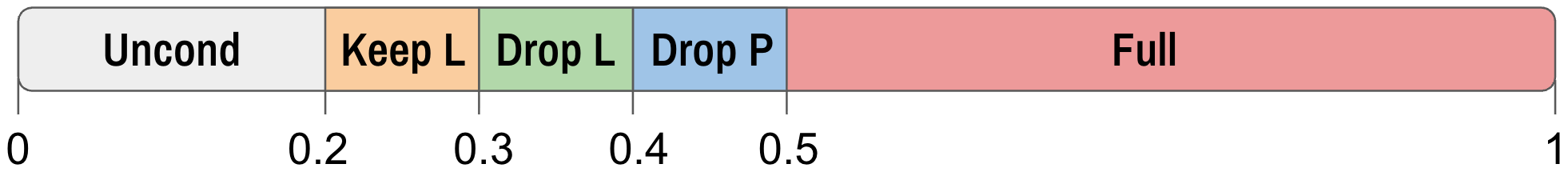}
\caption{Multi-level conditioning dropout schedule during flow training. Each mini-batch uses one mode per sample: unconditional (no conditioning), single-level variants (drop or keep one level), patch-level dropout (50\% of patches except for Level 0), or full hierarchical conditioning.}
\label{fig:dropout}
\end{figure}

Encoder embeddings at each hierarchy level are injected as conditioning signals into the corresponding U-Net layers.
To reduce conditioning dimensionality, we apply PCA with whitening, retaining 95\% of variance; this large compression still yields effective conditioning.
One important detail: whitening discards the mean, and the mean pitch carries musically meaningful information (the overall register of a passage) yet has insufficient variance to survive the 95\% threshold.
We therefore append the mean pitch explicitly as an additional conditioning dimension.

Training uses a ``Multi-Level Dropout'' scheme that serves as an extension to typical classifier-free guidance (CFG)~\cite{ho2022cfg} training: conditioning signals are randomly dropped out during training following the multi-level scheme in Figure~\ref{fig:dropout}, allowing unconditional, partially-conditioned, and fully-conditioned generation at inference by varying which level embeddings are provided.

\section{Results}

\subsection{Encoder Probes}

\begin{figure}[tbp]
\centering
\includegraphics[width=\columnwidth]{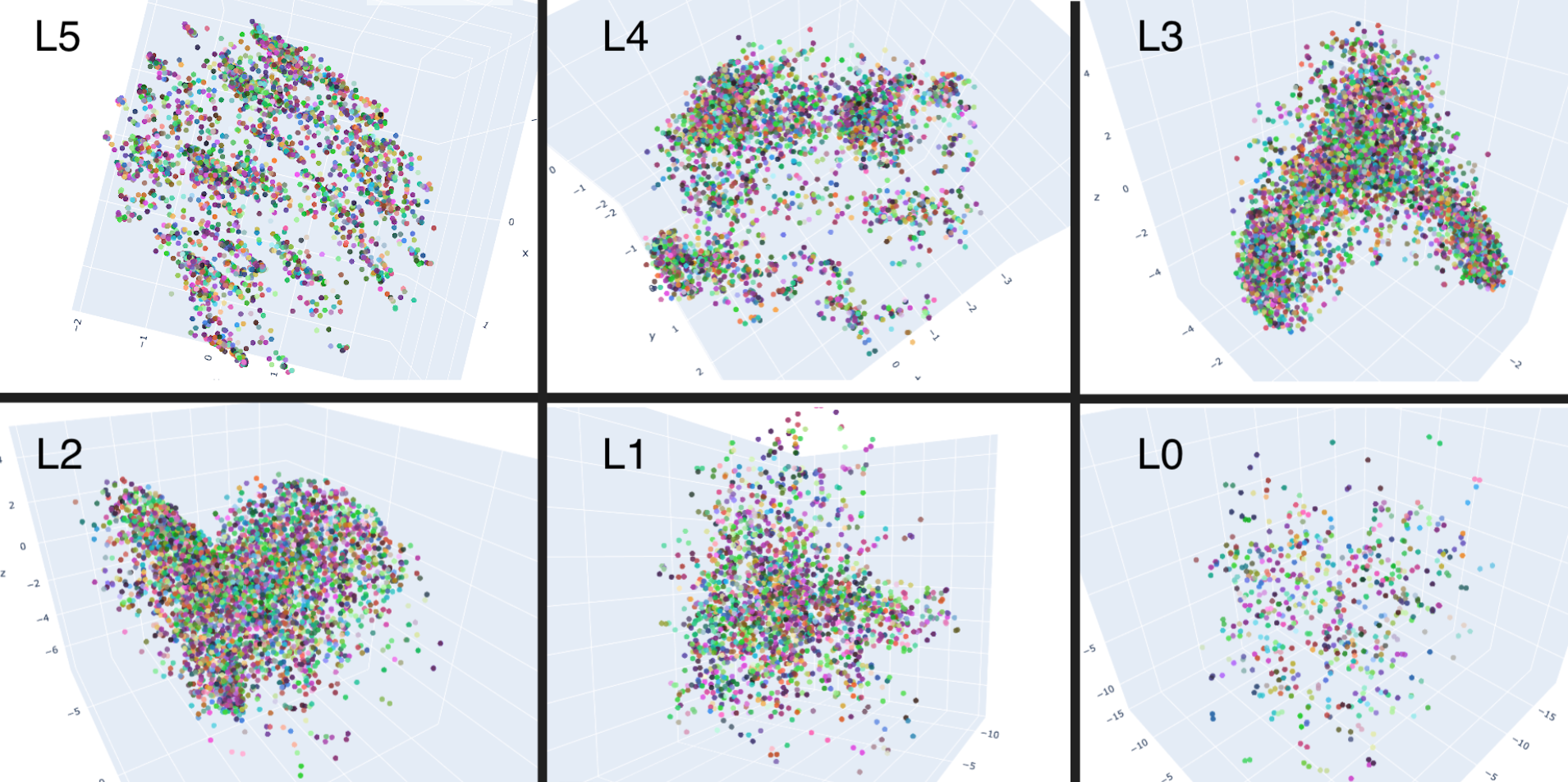}
\caption{PCA scatter plots of encoder embeddings at each hierarchy level (L5=finest, L0=coarsest). Finer levels show tighter, more separated clusters; coarser levels approach an isotropic Gaussian distribution.}
\label{fig:emb_pca}
\end{figure}

Figure~\ref{fig:emb_pca} shows PCA scatter plots of mean-pooled embeddings at each hierarchy level.
Finer levels (L5, L4) exhibit tight, well-separated clusters reflecting local note content, while coarser levels (L1, L0) show broader, more isotropic distributions consistent with SIGReg's Gaussian prior --- the coarser levels encode larger-scale musical context at the cost of local detail.

\subsubsection{Factorization}

\begin{figure}[tbp]
\centering
\includegraphics[width=\columnwidth]{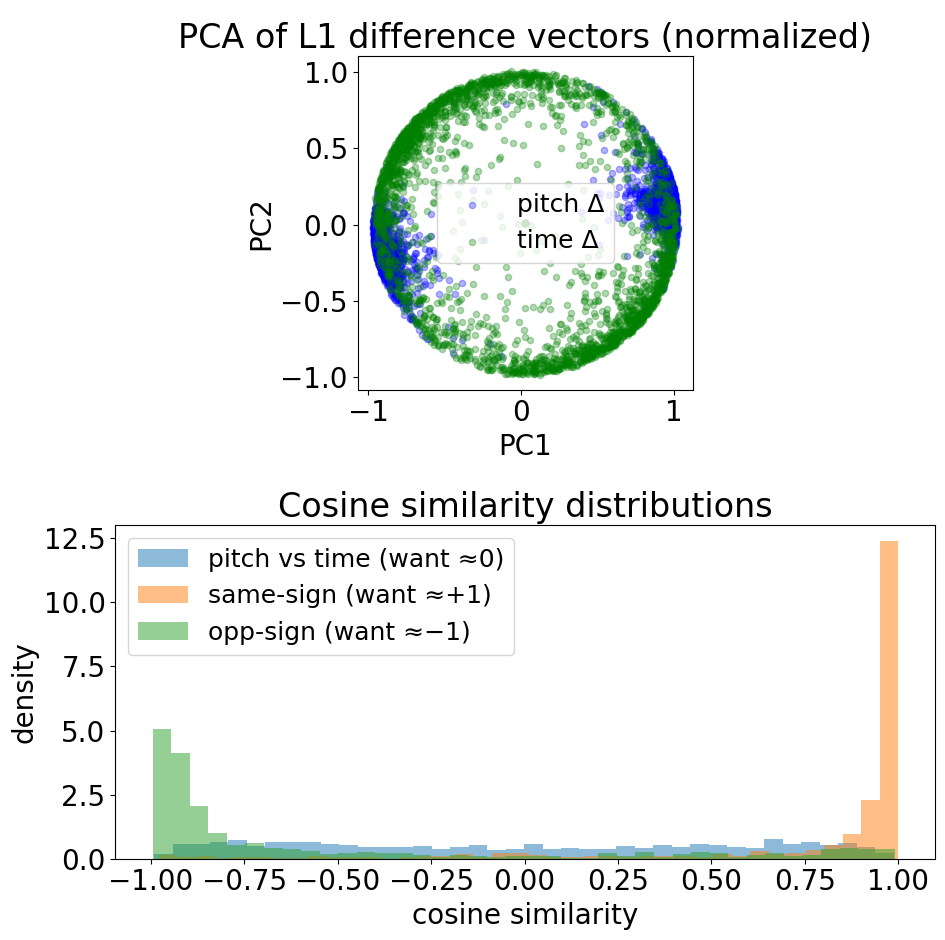}
\caption{Soft factorization results for level one. Left: PCA of normalized embedding difference vectors for the three target types shown in Figure~\ref{fig:softfact}, showing the expected geometric relationships. Right: histograms of cosine similarity distributions for each pair type, with means indicated by dashed lines.}
\label{fig:softfact_results}
\end{figure}

\vspace{-.4\baselineskip}
The factorization loss encourages pitch-shift and time-shift directions to be geometrically orthogonal in the embedding space. We verify this by computing embedding difference vectors for pairs of augmentations and measuring their cosine similarities. Table~\ref{tab:fact} shows that cross-type pairs (one pitch, one time shift) are nearly orthogonal ($|\cos| < 0.02$), same-type same-sign pairs are strongly parallel ($\cos \approx 0.78$), and same-type opposite-sign pairs show clear anti-parallel alignment ($\cos \approx -0.5$). The loss establishes these geometric relationships at L0--L2; L3 shows partial structure despite no factorization loss applied, suggesting information flow through the hierarchy. Figure~\ref{fig:softfact_results} visualizes the L1 geometry: PCA projections show pitch and time directions separating into distinct axes, and cosine histograms confirm the expected alignment patterns.

\begin{table}[tbp]
\centering
\caption{Cosine similarity of embedding difference vectors at each hierarchy level. Target values: Cross pairs = 0 (orthogonal), same-sign pairs = +1 (parallel), opposite-sign pairs = -1 (anti-parallel).}
\label{tab:fact}
\begin{tabulary}{\linewidth}{lcCC}
\toprule
Level & Cross & Parallel & Anti \\
\midrule
0 & -0.017 & 0.786 & -0.552 \\
1 & -0.014 & 0.793 & -0.549 \\
2 & -0.001 & 0.780 & -0.501 \\
3* & -0.014 & 0.727 & -0.354 \\
\bottomrule
\end{tabulary}
\end{table}
\vspace{-1em}

\vspace{.5\baselineskip}
\subsubsection{Equivariance}

\begin{figure}[tbp]
\centering
\includegraphics[width=\columnwidth]{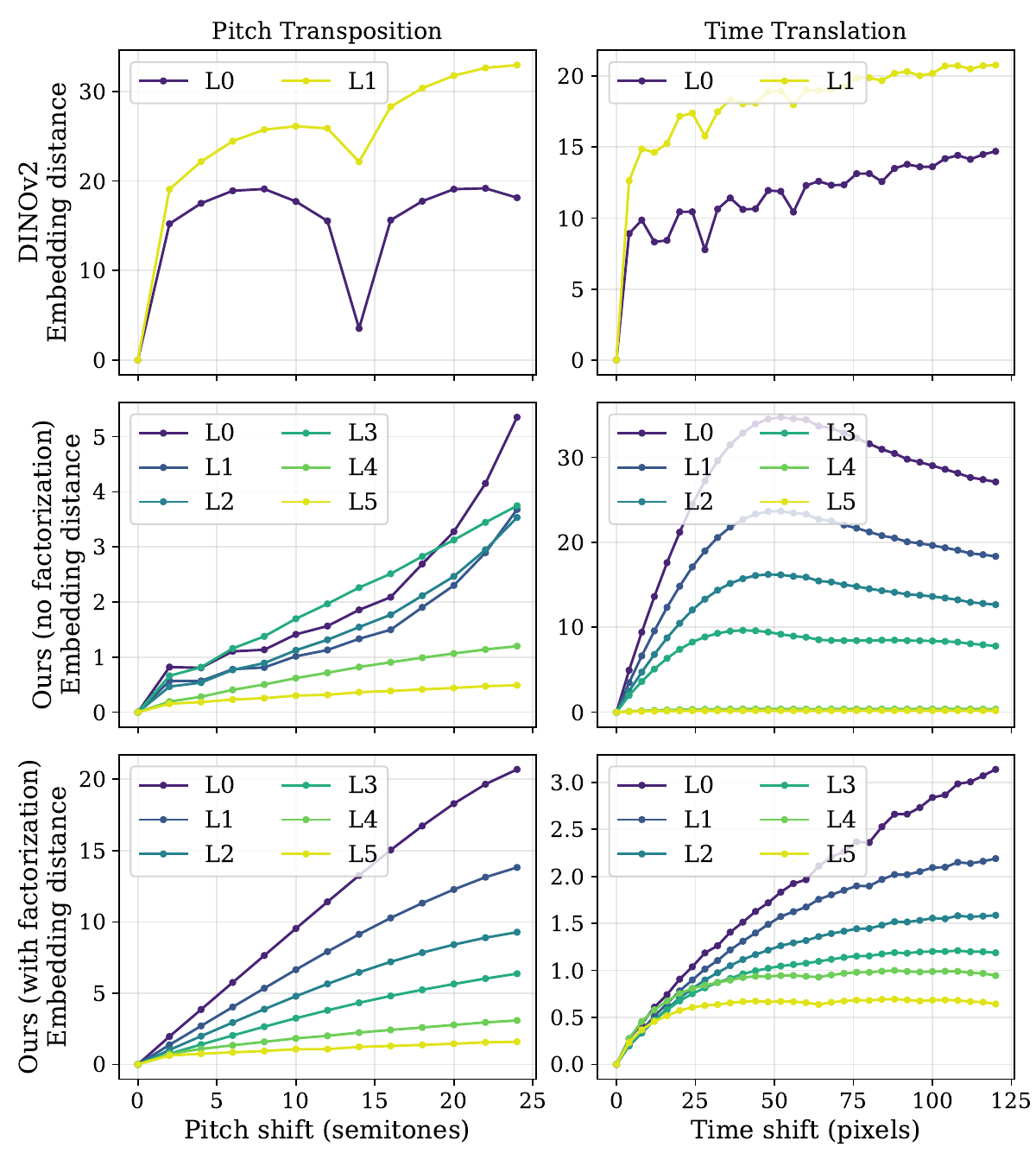}
\caption{Equivariance curves: embedding distance vs.\ pitch transposition (left) and time translation (right) at each hierarchy level, for DINOv2 (top), Our Model (middle), and Our model with factorization loss (bottom). Distances increase monotonically, extending beyond boundaries of small patches, but typically saturate after twenty pixels of time. The factorization loss establishes monotonicity for greater distances.}
\label{fig:equivariance}
\end{figure}

\vspace{-.4\baselineskip}
Figure~\ref{fig:equivariance} shows embedding distance as a function of pitch and time shift magnitude at each hierarchy level.
Distances increase monotonically with shift magnitude, including well beyond the nominal patch size at each level --- an effect we attribute to Swin's shifted-window attention allowing patches to attend across boundaries. While equivariance enforces geometric structure, the downstream benefits likely arise from the encoder learning musically meaningful pitch-interval and rhythmic-period relationships that transfer to tasks requiring harmonic and metric understanding, such as predicting 
which structures occur outside the patch size, or how far apart two passages are in time.

\subsubsection{Downstream task: emotion recognition}

\vspace{-.5\baselineskip}
We evaluate on EMOPIA~\cite{emopia}, a dataset of 1,071 piano clips labelled with four emotion quadrants (Russell's circumplex: arousal $\times$ valence).
Linear probes are trained on mean-pooled embeddings (4,000 crops) at each level and compared against Haar 
scattering~\cite{haar_scattering} and DINOv2~\cite{dinov2} baselines.
Table~\ref{tab:probes-emotion} shows the emotion results: our L3 embeddings
achieve the best scores on all three metrics, outperforming both baselines,
with performance falling off toward both the finest and coarsest levels.
Table~\ref{tab:probes-structure} reports complementary structure probes.

\begin{table}[tbp]
\centering
\caption{Emotion recognition probes (EMOPIA). Chance: 0.29 (4-class), 0.50 (Arousal/Valence). $\uparrow$ = higher better. Bold = best per metric.}
\label{tab:probes-emotion}
\begin{tabulary}{\linewidth}{lcCC}
\toprule
Model, Level & 4-class $\uparrow$ & Arousal $\uparrow$ & Valence $\uparrow$ \\
\midrule
\textit{Chance} & .290 & .500 & .500 \\
\midrule
Haar & .411 & .726 & .615 \\
\midrule
DINOv2,~L0 & .396 & .712 & .573 \\
DINOv2,~L1 & .391 & .697 & .584 \\
\midrule
MRJ-12,~L0 & .316 & .568 & .564 \\
MRJ-12,~L1 & .349 & .631 & .534 \\
MRJ-12,~L2 & .436 & .725 & .591 \\
MRJ-12,~L3 & \textbf{.488} & \textbf{.754} & \textbf{.640} \\
MRJ-12,~L4 & .462 & .737 & .593 \\
MRJ-12,~L5 & .425 & .736 & .571 \\
\bottomrule
\end{tabulary}
\end{table}

\begin{table}[tbp]
\centering
\caption{Structure probes. $\uparrow$ = higher better, $\downarrow$ = lower better. Bold = best per metric.}
\label{tab:probes-structure}
\begin{tabulary}{\linewidth}{lcCC}
\toprule
Model, Level & Density $R^2$ $\uparrow$ & Cross-song $\downarrow$ & Temp $R^2$ $\uparrow$ \\
\midrule
DINOv2,~L0 & .930 & .770 & .256 \\
DINOv2~L1 & .943 & .740 & \textbf{.307} \\
\midrule
MRJ-12,~L0 & .208 & .936 & .093 \\
MRJ-12,~L1 & .563 & .928 & .084 \\
MRJ-12,~L2 & .803 & .860 & .075 \\
MRJ-12,~L3 & .930 & .785 & .038 \\
MRJ-12,~L4 & \textbf{.991} & .682 & .157 \\
MRJ-12,~L5 & .984 & \textbf{.640} & .177 \\
\bottomrule
\end{tabulary}
\end{table}

Note density is strongly encoded at fine levels (L4/L5 $R^2 > 0.98$) but poorly at coarse levels, consistent with the PCA visualizations.
Cross-song similarity ratio decreases monotonically from L0 to L5, indicating that finer levels are increasingly discriminative between songs.
Temporal distance regression shows modest but consistent correlation at all levels; DINOv2 outperforms our encoder on this metric.

\subsection{Decoder Performance}

The decoder achieves high-fidelity reconstruction with an F1 score of approximately 0.995 on a held-out test set (Table~\ref{tab:decoder-f1}), confirming that the frozen encoder embeddings retain sufficient information for accurate piano roll reconstruction.
Extending the max time shift during encoder training can have a detrimental effect on reconstruction accuracy, likely because larger shifts encourage the encoder to focus more on coarse-level structure at the expense of fine-level detail; however, the factorization loss appears to mitigate this effect, as seen by the improved F1 of MRJ-48$\wedge$ compared to MRJ-48.


\begin{table}[tbp]
\centering
\caption{Decoder reconstruction performance, for our models trained with different maximum time shift $\Delta t_{max}$ and with/without the factorization loss, compared against a DINOv2 baseline. $\uparrow$ = higher better.  The numbers are close enough to be within the noise floor of training runs, yet the factorization seems to help, as seen by the slightly higher F1 of MRJ-48$\wedge$ compared to MRJ-48.}
\label{tab:decoder-f1}
\begin{tabulary}{\linewidth}{lcCL}
\toprule
Encoder & ${\Delta t}_max$ & Factorization & F1 $\uparrow$ \\
\midrule
MRJ-12 & 12 & no & 0.9955 \\
MRJ-48 & 48 & no & 0.981 \\
MRJ-48$\wedge$ & 48 & yes & 0.9953 \\
DINOv2 & n/a & n/a & 0.987 \\
\bottomrule
\end{tabulary}
\end{table}

\subsection{Conditional Generation}

We first attempted generation directly in representation space, but this did not produce satisfactory results.
Flowing in pixel space conditioned on encoder representations proved fast and highly effective.
PCA visualization of embeddings at each hierarchy level shows the expected structure: coarse levels cluster nearby crops tightly while fine levels separate by shift magnitude, with approximately isotropic distributions consistent with SIGReg's Gaussian prior.

\begin{figure}[tbp]
\centering
\includegraphics[width=\columnwidth]{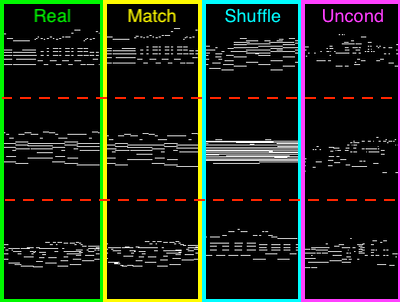}
\caption{Conditional generation results on POP909. Each row is a different excerpt. \textbf{Real}: ground truth piano roll. \textbf{Match}: generation conditioned on embeddings from that same excerpt (PCA 95\% variance + mean pitch). \textbf{Shuffle}: conditioned on embeddings from a randomly selected excerpt. \textbf{Uncond}: unconditional generation. Matched generations closely follow the rhythmic density and pitch register of the real excerpt; shuffled and unconditional samples are musically plausible but unrelated to the target.}
\label{fig:gen_columns}
\end{figure}

\begin{figure}[tbp]
\centering
\includegraphics[width=\columnwidth]{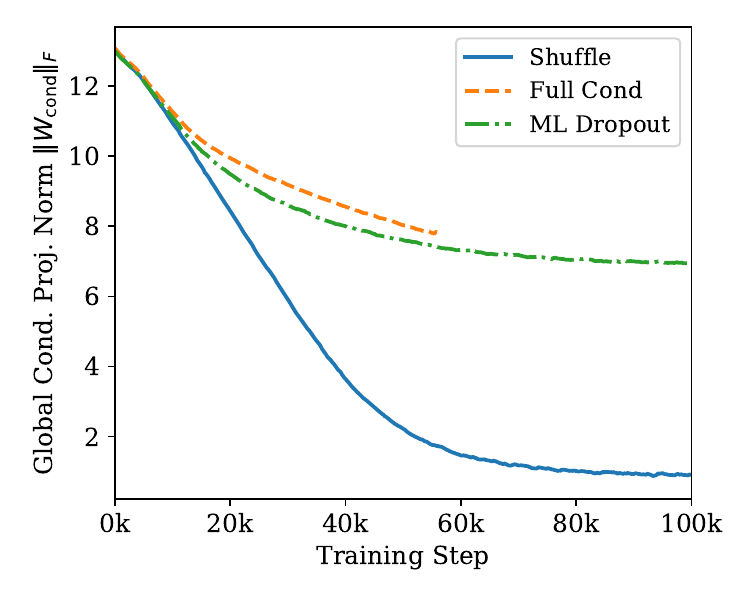}
\caption{Frobenius norm of the global conditioning projection $W_{\mathrm{cond}}$ over training steps. When conditioning is shuffled, the model learns to ignore it ($\|W_{\mathrm{cond}}\|_F \to 0$). Both full and multi-level dropout conditioning (Fig.~\ref{fig:dropout}) yield stable, nonzero weights, with dropout producing slightly lower values, suggesting marginally reduced reliance on the conditioning signal.}
\label{fig:cond_proj_norm}
\end{figure}

Figure~\ref{fig:gen_columns} shows real samples alongside matched, shuffled, and unconditional generations.
Matched generations closely follow the rhythmic and pitch character of the conditioning excerpt.
Shuffled conditioning (mismatched encoder embeddings) and unconditional generation produce plausible but unrelated piano rolls, confirming that the encoder embeddings carry musically meaningful information that the flow model successfully exploits.

As a further check that the trained flow genuinely uses the conditioning signal,
Figure~\ref{fig:cond_proj_norm} tracks the norm of the conditioning projection
during training: with shuffled (uninformative) conditioning the model learns to
ignore it, whereas full and multi-level dropout conditioning maintain stable,
nonzero weights.

\section{Conclusion}

We have shown that a self-supervised equivariance objective, combined with a Swin V2 encoder and LeJEPA, produces representations of symbolic music that are useful across multiple downstream tasks without any end-to-end training.
A decoder trained on frozen encoder embeddings achieves F1 $\approx$ 0.995, confirming that the SSL objective alone forces the encoder to retain sufficient information for high-fidelity reconstruction.
The soft factorization loss, which encourages pitch and time directions to be orthogonal in latent space, improves equivariance and resists saturation of embedding distances at large shifts.
On the EMOPIA emotion classification task, a linear probe on our embeddings outperforms a Haar scattering baseline, suggesting that musically relevant information is accessible without fine-tuning.
Conditioning a flow matching generative model on PCA-reduced encoder embeddings produces outputs that closely track the pitch register and rhythmic density of the conditioning excerpt, and unconditional or mismatched conditioning yields musically plausible but unrelated output---demonstrating that the representations carry generatively useful structure.

Most of the representational capacity resides at the finer levels of the hierarchy; the coarser levels do not yet capture obviously interpretable musical abstractions.
Getting more musically meaningful structure into those levels---whether through modified objectives, explicit labels, or architectural changes---is an open problem.
More broadly, extending this approach to audio spectrograms, scaling to larger datasets, and exploring richer downstream tasks such as arrangement or accompaniment generation are natural next steps.

\section{Acknowledgements}

Thanks TwinOS and Razer Corporation for making available the Blade 16'' and Blade 18'' GPU laptops 
used for some of the computations in this paper. Thanks to Vincent Lostanlen for the suggestion of 
the Haar wavelet baseline. 

\section{AI Usage Statement}

A large language model (Claude, Anthropic)
was used as a coding assistant for debugging, visualization,
and test generation; as an experiment management assistant
for launching jobs and tracking results; and as an interactive
writing assistant in editing this paper, including portions of
the introduction and related work. All scientific hypotheses,
experimental designs, architectural decisions, and conclusions are the author’s own, and all content, references, and
claims were verified by the author.

\small

\bibliography{midi_rae_refs}

\end{document}